\documentclass{emulateapj}
\usepackage{apjfonts}

\newcommand{\timav}{\langle \dot M \rangle}
\newcommand{\be}{\begin{eqnarray}}
\newcommand{\ee}{\end{eqnarray}}

\newcommand{\pder}[2]{\frac{\partial #1}{\partial #2}}
\shorttitle{CNe and the CV Population}

\begin{document}

\submitted{Accepted to the ApJ}
\title{Classical Novae as a Probe of the Cataclysmic
Variable Population} 

\author{Dean M. Townsley\altaffilmark{1}}
\affil{Department of Physics,\\
Broida Hall, University of California, Santa Barbara, CA 93106}
\altaffiltext{1}{Current Address: Department of Astronomy and Astrophysics,\\
5640 South Ellis Avenue, University of Chicago, Chicago, IL 60637;
townsley@uchicago.edu}

\and

\author{Lars Bildsten}
\affil{Kavli Institute for Theoretical Physics and Department of Physics,\\
Kohn Hall, University of California, Santa Barbara, CA 93106;
bildsten@kitp.ucsb.edu}

\begin{abstract}

Classical Novae (CNe) are the brightest manifestation of mass transfer
onto a white dwarf in a cataclysmic variable (CV). As such, they are
probes of the mass transfer rate, $\dot M$, and WD mass, $M_{\rm WD}$, in these
interacting binaries. Our calculations of the dependence of the CN
ignition mass, $M_{\rm ign}$, on $\dot M$ and $M_{\rm WD}$ yields the
recurrence times of these explosions. We show that the observed CNe
orbital period distribution is consistent with the interrupted
magnetic braking evolutionary scenario, where at orbital periods
$P_{\rm orb}> 3\ {\rm hr}$ mass transfer is driven by angular momentum
loss via a wind from the companion star and at $P_{\rm orb}<3\ {\rm
hr}$ by gravitational radiation. About 50\% of CNe occur in binaries
accreting at $\dot M\simeq 10^{-9}M_\odot~{\rm yr}^{-1}$ with $P_{\rm
orb}=3-4 \ {\rm hr}$, with the remaining 50\% split evenly between
$P_{\rm orb}$ longer (higher $\dot M$) and shorter (lower $\dot M$)
than this. This resolution of the relative contribution to the CN rate
from different CVs tells us that $3(9)\times 10^5$ CVs with WD mass
$1.0(0.6)M_\odot$ are needed to produce one CN per year. In addition,
one CN per year requires a CV birthrate of $1(2)\times 10^{-4} \ {\rm
yr^{-1}}$, and likely ejects mass into the ISM at a rate $\dot
M=3(9)\times 10^{-5}M_\odot\ {\rm yr^{-1}}$.  Using the K-band
specific CN rate measured in external galaxies, we find a CV birthrate
of $2(4)\times 10^{-4}$ yr$^{-1}$ per $10^{10}L_{\odot,K}$, very similar to
the luminosity specific Type Ia supernova rate in elliptical galaxies.
Likewise, we predict that there should be 60-180 CVs for every
$10^6L_{\odot,K}$ in an old stellar population.  The population of
X-ray identified CVs in the globular cluster 47 Tuc is similar to this
number, showing no overabundance relative to the field.

 The observed CN $P_{\rm orb}$ distribution also contains evidence for
a CV population which has no period gap. These are likely systems with
a strongly magnetic WD (Polars) in which it has been suggested that
the field interferes with the wind of the companion, limiting angular
momentum losses to those of gravitational radiation and eliminating
the period gap.  With this reduced $\dot M$, Polars evolve more slowly
than systems which undergo magnetic braking.  Using a two-component
steady state model of CV evolution we show that the fraction of CVs
which are magnetic (22\%) implies a birthrate of 8\% relative to
non-magnetic CVs, similar to the fraction of strongly magnetic field
WDs.

\end{abstract}


%%%%%%%%%%%%%%%%%%%%%%%%%%%%%%%%%%%%%%%%%%%%%%%%%%%%%%%%%%%%%%%%%%%%%%%%
\section{Introduction}

A Classical Nova (CN) outburst is the result of a thermonuclear runaway in
the hydrogen-rich accreted layers on a white dwarf (WD) in a mass
transferring cataclysmic variable (CV) binary (see \citealt{Shar89} for a
review).  The number of CNe with measured orbital periods, $P_{\rm orb}$, has
been increasing; \citet{DiazBruc97} made a first comparison to the CN period
distribution with 30 objects, and recently \citet{Warn02} listed $P_{\rm
orb}$ for 50 CNe.  These range from 1.4 hours to more than 16 hours, with
most below 5 hours.  Cataclysmic variables (CVs; \citealt{Warn95}) consist of
a WD primary accreting mass from a normal low-mass star that is filling its
Roche lobe, and spend most of their life accreting at rates $\dot M\approx
10^{-9}$--$10^{-11}M_\odot$ yr$^{-1}$ (e.g. \citealt*{Howeetal01}, herafter
HNR).  CVs are formed when the companion (of mass $M_2$) to a WD (made during
a common envelope event) comes into contact with its Roche lobe as a result
of angular momentum losses.  The CV evolves toward shorter orbital periods as
the binary loses angular momentum at a rate $\dot J$ and the companion
transfers mass to the WD at a rate $\dot M$.  The orbital period distribution
of CN, a direct reflection of the CN frequency, provides a diagnostic of
$\dot M$ in CVs over the time to accumulate an unstable layer, $10^{5{-}7}$
years.

It is generally thought that the paucity of CVs with 2 hr $\lesssim P_{\rm
orb}\lesssim$ 3 hr (e.g.~\citealt{Shaf92}), the so-called period gap, is due
to switching between a high $\dot J$ state above 3 hours to a comparatively
low $\dot J$ state below 3 hours (e.g.~HNR).  Upon such a sudden reduction in
$\dot J$, the requisite reduction in $\dot M$ causes the companion, which has
been slightly out of thermal equilibrium due to the high mass loss rate, to
contract and fall inside the Roche lobe, halting mass transfer.  Eventually
as $P_{\rm orb}$ continues to decrease due to angular momentum loss, Roche
lobe contact is reestablished.  The simplest paradigm of CV evolution holds
that $\dot J$ is determined by the WD mass $M_{\rm WD}$, $M_2$ and $P_{\rm
orb}$.  Interrupted magnetic braking (IMB) has arisen as a good candidate for
$\dot J$ prescriptions (see e.g.\ \citealt{PaczSien83}; \citealt{Rappetal83};
\citealt{SpruRitt83}; \citealt{Hameetal88}; \citealt{Kolb93}; HNR).  In the
IMB scenario, $\dot J$ is initally driven by angular momentum lost in the
stellar wind of the secondary star \citep{VerbZwaa81} for $P_{\rm orb}\gtrsim
3$ hours.  The magnetic braking is interrupted (potentially by the loss of
the radiative core), and $\dot J$ falls to that carried away by gravitational
radiation (see e.g.\ \citealt{Faul71}; \citealt{PaczSien81};
\citealt{Rappetal82}).  IMB explains the period gap of the CV population and
the predicted $\timav$ agrees with observations of the WD surface
temperatures \citep{TownBild03}.  This agreement is somewhat surprising
because modern measurements of $\dot J$ for isolated low mass stars do not
appear to follow the law used in the IMB scenario \citep{Andretal03}.  IMB
also makes a very important but unconfirmed prediction: that CVs evolve
across the period gap.

For a given composition of accreted material, there are three independent
parameters that set $M_{\rm ign}$, and thus the CN rate: $\dot M$, $T_c$, and
$M_{\rm WD}$.  Many previous studies (e.g.~\citealt{TrurLivi86};
\citealt{Rittetal91}) focused only on the dependence of the CN frequency on
$M_{\rm WD}$.  Here we focus on $T_c$'s dependence on $\dot M$.
\citet{TownBild04} (hereafter TB04) showed
that, after extended accretion and many CN cycles, $T_c$ is set by $\dot M$
because of accretion's impact on the thermal state of the WD core during the
long interval between CNe.  Given any $\dot M(P_{\rm orb})$ relation, TB04's
calculation of $T_c$ enables the first consistent prediction of how the CN
rate varies with orbital period.  Using $\dot M$ dependent $M_{\rm ign}$
values calculated at several $T_c$ values \citep{PriaKove95},
\citet{Nelsetal04} showed clearly that this knowledge of $T_c$ is essential
for a conclusive prediction of the $P_{\rm orb}$ distribution.  In Section
\ref{sec:CN} we briefly review the work of TB04, and in Section
\ref{sec:perioddist} we apply it to calculate the CN period distribution and
compare to that observed.  We find that IMB is remarkably consistent with the
observed CN frequency, yielding a firm relationship between observed CN
frequency and the CV population.  Confirming this scenario finally
allows us to use the observed CN rate to infer the number of underlying CVs.
This impacts searches for CVs in the field, as well as in
globular clusters, as we note in Section \ref{sec:discussion}.

%%%%%%%%%%%%%%%%%%%%%%%%%%%%%%%%%%%%%%%%%%%%%%%%%%%%%%%%%%%%%%%%%%
\section{Classical Nova Ignition Masses}
\label{sec:CN}

IMB predicts that typical accretion rates near $P_{\rm orb}=3$-4 hours are
$\dot M\simeq 10^{-9}M_\odot\ {\rm yr^{-1}}$, and decrease to $\dot M \simeq
5\times 10^{-11} M_\odot\ {\rm yr^{-1}}$ below 2 hours.  This factor of 20
drop in $\dot M$ across the period gap requires that the $\dot M$ dependence
of $M_{\rm ign}$ be taken into account when explaining the orbital period
distribution of CN.  These two $\dot M$'s lead to different ignition modes.
For the higher $\dot M$, TB04 found that the CN is triggered by trace $^3$He
in the accreted material \citep{Shar80}, leading to a CNO burning runaway in
a radiative layer.  At the lower $\dot M$, however, the CN outburst is
triggered by the $pp$ chain when the base of the accreted shell is
conductive.  This wide range of conditions means that the prior
prescriptions of ignition, such as a single unique pressure
(\citealt{Gehretal98}; \citealt{Warn02}) or taking $M_{\rm ign}$ independent
of $\dot M$ (\citealt{Rittetal91}; \citealt{DiazBruc97}), are wholly
inadequate.

By considering the release of energy due to compression of the accreted
layers, TB04 calculated the rate of heat loss or gain of the WD core.  This
rate varies as the accreted layer builds up, and by taking the average loss
or gain to be zero TB04 found an equilibrium, $T_{c,\rm eq}$, which the WD will
approach under prolonged accretion.  For $P_{\rm orb}\gtrsim 3$ hours, the
magnetic braking phase, the WD has only marginally enough
time, $\sim 10^7$ years, to reach $T_{c,\rm eq}$.  However, at these
accretion rates, $M_{\rm ign}$ is insensitive to $T_c$ if it is lower than
$T_{c,\rm eq}$ because the ignition occurs in a layer which is not
conductively coupled to the core.  Hence the temperature is set by the
response of the envelope to the energy being released by compression (TB04).
At lower $\dot M$, $T_c$ determines $M_{\rm
ign}$,  but the low $\dot M$ state for CVs is long-lived ($>10^9$ years),
allowing the WD core to reach $T_{c,\rm eq}$.  Thus, for CV evolution, use
of $T_{c,\rm eq}$ provides $M_{\rm ign}$'s which are self-consistent and
independent of assumptions about the WD thermal state. This is in contrast to
all prior $M_{\rm ign}$ calculations (see e.g.\ \citealt{Fuji82};
\citealt{MacD84}; \citealt{PriaKove95}).  The $T_c$ values found by TB04 are
$\lesssim 10^{7}$ K for $M_{\rm WD}<1.2M_\odot$, lower than the lowest $T_c$
considered by \citet{PriaKove95} in their broad parameter survey.

Models of donor evolution in CVs \citep{DAntMazz82,IbenTutu84,McDeTaam89}
show that the $^3$He abundance reaches mass fractions of $X_{^3\rm He}\simeq
0.001$ when the orbital period is 3-4 hours (the companion mass is below
$\simeq 0.4M_\odot$).  This is when the star has lost enough mass and the
convection layer has become deep enough that the partially-fused material in
the stellar interior is uncovered.  Precisely when this occurs and the
$X_{^3\rm He}$ reached is sensitive to the donor's initial mass, but
generally does not exceed 0.003.  We are most interested in CN rates for
$P_{\rm orb}\le 4$ hours, so we include a fraction $X_{^3\rm He}=0.001$ in
our ignition models.

\begin{figure}
\plotone{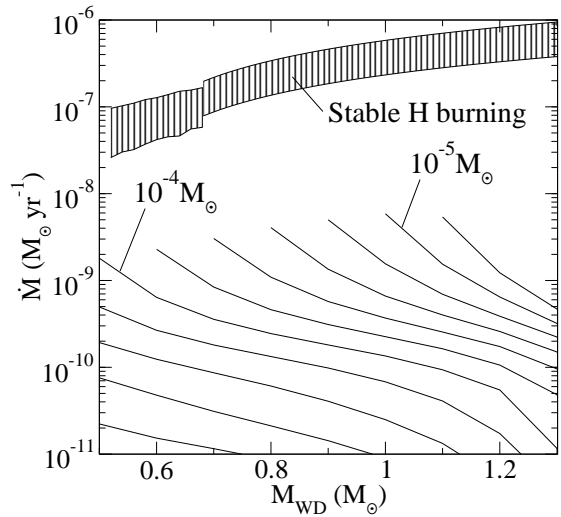}
\caption{
\label{fig:Mign}
Classical Nova ignition masses, $M_{\rm ign}$, for WDs which have reached
their equilibrium core temperature, as determined by TB04.
Contours are evenly spaced in $\log(M_{\rm ign})$, labels indicate $M_{\rm
ign}$ in $M_\odot$.  The vertically hatched region indicates where steady
burning of H is expected \citep{Pieretal00,Nomo82}.  At higher $\dot M$,
envelope buildup and expansion into a giant is expected.
}
\end{figure}
The $M_{\rm ign}$ results of TB04 are shown in Figure \ref{fig:Mign}.
Compared to previous ignition mass calculations, TB04 found larger $M_{\rm
ign}$ at $\dot M \lesssim 10^{-10} M_\odot\ {\rm yr^{-1}}$ due to the lower
$T_c$, and similar $M_{\rm ign}$ at $M_{\rm ign} \gtrsim 10^{-9}M_\odot~{\rm
yr^{-1}}$ due to offseting effects of the lower $T_c$ and the inclusion of
$^3$He leading to an earlier trigger.  TB04 found that for the few systems
which have a known ejected mass and $P_{\rm orb}$ (giving some indication of
$\dot M$), the ejected masses are similar to $M_{\rm ign}$, implying that
each CN ejects approximately the amount of material accreted since the last
event.  TB04's results do not show the trivial $M_{\rm ign}\propto R^4/M_{\rm WD}$
scaling assumed for a unique ignition pressure (see their discussion).  An
ignition pressure of $p=2\times 10^{19}$ dyne cm$^{-2}$ would have $10^{-5}$,
$10^{-4}$ and $10^{-3}M_\odot$ contours which are vertical lines at $M_{\rm
WD}=1.3$,
1.0 and $0.58M_\odot$.  (This ignition pressure is used by \citet{Gehretal98}
in their discussion of the nucleosynthetic impact of nova ejecta.)  As we
will see, most CNe have $\dot M\sim 10^{-9}M_\odot~{\rm yr^{-1}}$, and thus
lower $M_{\rm ign}$ than this prescription implies.

Also shown in Figure \ref{fig:Mign} is the region (hatched) where steady
burning is expected.  For $0.52\le M_{\rm WD}/M_\odot \le 0.68$, the
calculations of \citet{Pieretal00} are shown. For $M_{\rm WD}\ge 0.6M_\odot$
the upper bound is found from the core mass-luminosity relation of
\citet{Pacz70}, and the lower bound is taken to be 0.4 of this
\citep{Nomo82}.  At higher $\dot M$, envelope buildup and expansion into a
giant is expected.

%%%%%%%%%%%%%%%%%%%%%%%%%%%%%%%%%%%%%%%%%%%%%%%%%%%%%%%%%%%%%%%%%%
\section{The Classical Novae Orbital Period Distribution}
\label{sec:perioddist}

\begin{figure}
%\epsscale{0.7}
\plotone{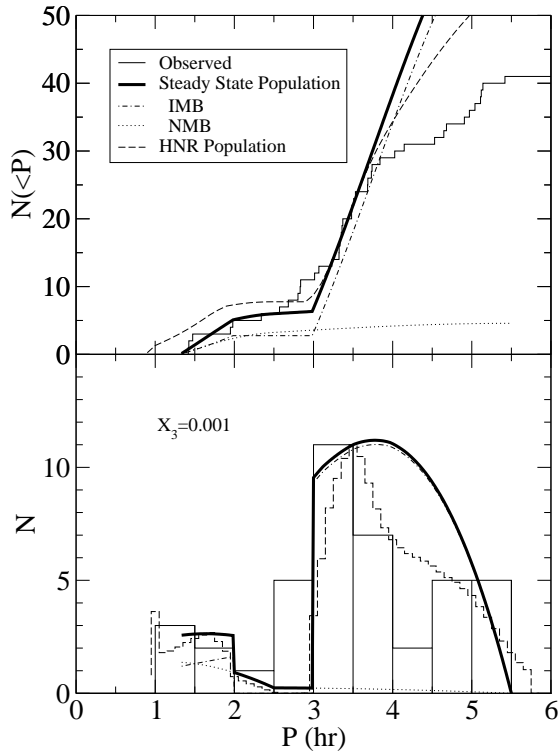}
\caption{\label{fig:cumanddist}
The cumulative (top panel) and normal (bottom panel) distribution of
classical novae orbital periods.
The thin solid lines show the observed distribution, with
the 9 systems with $P_{\rm orb}>6$ hours not shown.  The steady state
population described in the text ({\it thick solid line}), consists of an
equal number of CVs which undergo an interrupted magnetic braking episode
(IMB population, {\it dot-dashed line}) and which undergo no magnetic braking
(NMB population, {\it dotted line}). This combined population reproduces the
major features of the cumulative distribution for $P_{\rm orb}\lesssim 4$
hours.  A full population synthesis (HNR, {\it dashed line})  including only
objects which undergo a magnetic braking phase displays similar features,
matching the observed distribution remarkably well.  A mass fraction
$X_{^3\rm He}=0.001$ is used for all.
}
\end{figure}

The observed distribution of CN orbital periods \citep{Warn02} is shown in
Figure \ref{fig:cumanddist} by the thin solid line.  These data differ only
very slightly from the collection in \citet{RittKolb03}.
We have neglected any selection effects, as they have not been
satisfactorily quantified to date.  These are complex, so we will not attempt
a conclusive discussion, only mention two important ones, CN outburst
brightness and remnant brightness.  The tables in \citet{PriaKove95}, for the
coldest $T_c$ and $M_{\rm WD} =1 M_\odot$, imply that CN outbursts for
systems with $\dot M\sim 10^{-10}M_\odot\ {\rm yr^{-1}}$ are 3-4 times as
bright as those for systems with $\dot M \sim 10^{-8}M_\odot$ yr$^{-1}$.
However, the spatial distribution in the Galaxy and the effects of absorption
must be considered in order to judge how the observed population reflects
this brightness variation. (See \citealt{Rittetal91} for some discussion.)
It is likely that remnant brightness depends on $\dot M$, and thus $P_{\rm
orb}$, introducing additional selection effects in remnant recovery and
characterization.

\subsection{Using Previous Population Synthesis Calculations}

In order to calculate the CN frequency we must first use the IMB
prescriptions to obtain $\dot M$ from $P_{\rm orb}$.
Since the secondary fills its Roche Lobe, $P_{\rm orb}$ is directly
related to $M_2$ by its mass-radius relation.  We enforce
$R_2 = 0.46a[M_2/(M_{\rm WD}+M_2)]^{1/3}$, so that $P_{\rm orb} = 9{\rm\ hr}\
(M_2/M_\odot)^{-1/2} (R_2/R_\odot)^{3/2}$.
To reproduce IMB for $P_{\rm orb}> 3$ hours we use the magnetic braking law
from HNR (see also \citealt*{Rappetal83})
\begin{equation}
\label{eq:mblaw}
%\dot J_{\rm mb} = -3.8\times 10^{-30} M_2 R_\odot^4(R_2/R_\odot)^3
%\omega^3\ {\rm dyne\  cm},
\dot J_{\rm mb} = -9.4\times 10^{38}\ {\rm ergs}
\left(\frac{M_2}{M_\odot}\right)
\left(\frac{R_2}{R_\odot}\right)^3
\left(\frac{P_{\rm orb}}{\rm hr}\right)^{-3}.
\end{equation}
When the binary is evolving under the influence of magnetic braking, the
$\dot M_2$ is large enough that the secondary is out of thermal equilibrium
and therefore has a larger radius than a main sequence star of the same mass
(the effect that leads to the period gap).
In order to account for this we use the $M_2$-$R_2$ relation found by
HNR using a bipolytrope model for the secondary and evolving under
equation (\ref{eq:mblaw}),
$R_2/R_\odot = 0.81(M_2/M_\odot)^{0.67}$.  For $P_{\rm orb}<3$ hours, we apply
gravitational radiation losses given by
\begin{eqnarray}
\label{eq:grlaw}
\dot J_{\rm gr} &=& -\frac{32GQ^2\omega^5}{5c^5}\nonumber\\
&=&-2.7\times 10^{37}{\ \rm ergs}
\left(\frac{a}{R_\odot}\right)^4
\left(\frac{M_{\rm WD}M_2}{M_tM_\odot}\right)^2
\left(\frac{P_{\rm orb}}{\rm hr}\right)^{-5},
\end{eqnarray}
where $Q$ is the
moment of the binary about the orbital center,
$\omega =2\pi /P_{\rm orb}$, $a$ is the orbital separation and $M_t=M_{\rm
WD}+M_2$.
We use $R_2/R_\odot =
0.71(M_2/M_\odot)^{0.77}$ for a star in thermal equilibrium also from HNR.
These $M_2$-$R_2$ relations give a period gap (transition from the
former to the latter) at 2-3 hours when the donor is $0.21M_\odot$, as found
for this ``standard model'' (Model A) in HNR.

In addition to $M_{\rm ign}$ and the IMB prescriptions we also require a CV
population model to construct the CN orbital period distribution.  Detailed
Monte Carlo CV population models are explored by HNR and by
\citet{Nelsetal04}, in the same context as here, but without the TB04
predictions of $T_c$ from $\dot M$.  HNR calculated the number of CVs per
period interval, which we denote $n_P$ so that $n_PdP$ is the number with
orbital periods between $P$ and $P+dP$.  We use a single characteristic mass
in lieu of a detailed $M_{\rm WD}$ distribution.  For the HNR population we
use $M_{\rm WD}=0.72M_\odot$, the average CN $M_{\rm WD}$ for their
populations, and $X_{^3\rm He}=0.001$.

Using the above relationships to obtain $\dot M$, $n_P$ from HNR,
and the $M_{\rm ign}$ from
TB04, the resulting CN distribution is shown by the dashed line in Figure
\ref{fig:cumanddist}, and is normalized to match at $P_{\rm
orb}=3.5$ hours.  The most noticeable difference between this and our
steady-state model presented below (thick solid line) is how the CV
population based on HNR drops off at long $P_{\rm orb}$ due to the CVs being
born throughout the 3.5-6 hour range.  The exact form of this turnover
depends on the common envelope phase \citep{Nelsetal04} through the orbital
period distribution of the common envelope products.  By construction, the
HNR distribution lacks systems in the period gap.  Every system which comes
into contact with a companion that has a radiative core ($P_{\rm orb} \gtrsim
2.4$ hours) goes through a MB phase.  This distribution matches quite well in
both the 1-2 hour and 3-4 hour regions, keeping in mind the $\sqrt{N}$ noise
present in the observed distribution.  The HNR population extends down to 1
hour due to the famous discrepancy in the period minimum between observation
and theory \citep{KolbBara99}.

\subsection{Simpler Population Calculations} 

In order to understand how the $P_{\rm orb}$ distribution depends separately
on the distribution of $\dot M$ and the CV birth distribution, we
also construct a steady-state population with a specified
birth distribution.  CVs are born (i.e. come into contact and initiate mass
transfer for the first time) at a variety of orbital periods determined by
the distribution of $M_{\rm WD}$ and $M_2$ after the common envelope event.  For
a population of $n_P$ CVs per period interval with characteristic mass
$M_{\rm WD}$, conservation implies
\begin{equation}
\label{eq:cons}
\pder{n_P}{t} = \pder{(n_P\dot P_{\rm orb})}{ P_{\rm orb}} + b_P,
\end{equation}
where $b_P$ is the CV birth rate per period interval, i.e. the number of CVs
per year which commence accretion for the first time between $P$ and $P+dP$.
We assume that all matter accreted from the companion is ejected in the CN,
consistent with current observations (TB04), so that $M_{\rm WD}$ is
constant.  We also assume that the CV birthrate has been constant over the
last 4 Gyrs, the time it takes a CV to reach the period minimum at $P_{\rm
orb}\simeq 80$ minutes (HNR).  Beyond this point the companion becomes
essentially fixed in radius, so that the binary expands and $\dot M$ drops
quickly.  These assumptions give a steady-state ($\partial n_P/\partial t=0$)
CV population whose number density per period interval from equation
(\ref{eq:cons}) is
\begin{equation}
\label{eq:npint}
n_P = \frac{1}{|\dot P_{\rm orb}|}\int_{\infty}^{P_{\rm orb}} b_PdP,
\end{equation}
where we assume no systems at large $P_{\rm orb}$.

As noted above, the secondary's mass-radius relation is well characterized by
a power law, and so for Roche Lobe filling systems $\alpha \equiv d \ln
P_{\rm orb}/d \ln M_2$ is approximately a constant, typically between 0.5 and
1.  Thus $\dot P_{\rm orb}=-\dot M\alpha P_{\rm orb}/M_2$ and the CN
frequency per period interval is
\begin{equation}
\label{eq:nup}
\nu_P \equiv \frac{\dot M}{M_{\rm ign}}n_P
\propto \frac{M_2/P_{\rm orb}}{M_{\rm ign}}.
\end{equation}
The direct dependence on $\dot M$ cancels for this observable.  By far the
most important part of the remaining $\dot M$ dependence is in $M_{\rm ign}$.
Evaluating across the period gap, $M_2$ is constant, while $P_{\rm orb}$
changes by 50\% due to bloating of the companion above the gap.  On the other
hand $\dot M$ is expected to change from $10^{-9}M_\odot\ {\rm yr^{-1}}$ to
$\simeq 5\times 10^{-11}M_\odot\ {\rm yr^{-1}}$, causing $M_{\rm ign}$ (from
Figure \ref{fig:Mign}) to increase from $\simeq10^{-5}M_\odot$  to $\gtrsim
10^{-4}M_\odot$, a factor of $>10$.  Thus the CN orbital period distribution
directly reflects $\dot M$ through its effect on $M_{\rm ign}$.  With the
consistent calculation of $M_{\rm ign}$ from $\dot M$ provided by TB04, we
can probe the $\dot M$-$P_{\rm orb}$ relation directly by comparing the
steady state population to observations.  Evidence for evolution across the
gap is then contained in the degree to which a single steady state
distribution can reproduce the observations.

We use a binary with $M_{\rm WD} = 1.0M_\odot$ as representative, since the
actual mass distribution is very sensitive to the common envelope
prescription \citep{Nelsetal04}.  Also, we set $X_{^3\rm He}=0.001$ because
this is the approximate level expected in the accreted matter (See section
\ref{sec:CN}).  In order to account for the possibility of systems which do
not undergo magnetic braking, we use three populations.  The first follows
the IMB scenario (the IMB population), while the others only have $\dot
J_{\rm gr}$ applied (the NMB [No Magnetic Braking] populations).  In the NMB
case the secondary mass-radius relation is that of an equilibrium main
sequence star from HNR.

Though the detailed form of the birth distribution above the gap is sensitive
to the CE prescription (HNR; \citealt{Nelsetal04}), we employ a constant
$b_P$ between 3 and 5.5 hours for the IMB population.  This capures the
contrast in classical nova rate across the gap without depending explicitly
on a particular CE prescription.  CVs born with $P_{\rm orb}\lesssim 2.5$
hours do not undergo magnetic braking in the IMB scenario.  We call this
population the NMBb (NMB below gap) population, and it has $b_P$ constant
between 1.3 and 2.5 hours.  A second NMBa (NMB all periods) population
represents systems which do not undergo magnetic braking regardless of their
birth $P_{\rm orb}$ (see discussion of magnetic systems below).  This
population has a constant birth function between 1.3 and 5.5 hours.  Note
that, unlike the HNR population, we directly specify the minimum $P_{\rm
orb}$.  These birthrates can then be used in equation (\ref{eq:npint}) and
(\ref{eq:nup}), and the relative birth fractions for each population are
given by $b_P(P_{\rm max}-P_{\rm min})$.

This steady state population is shown as the thick solid line in Figure
\ref{fig:cumanddist}, arbitrarily normalized to match the observed
distribution (thin solid line) at 3.5 hours.  The relative birth fractions
for the curve shown are NMBa: 8\%, IMB: 46\%, NMBb: 46\%.
This model reproduces
the major features of the observed population below 4 hours, including the
slope in the cumulative distribution between 3 and 4 hours.  The dependence
on the common envelope prescription and binary population parameters is
essentially contained in the average $M_{\rm WD}$, to which the shape of the
curve is insensitive. The effect of a 30\% change in $M_{\rm WD}$ is small
compared to the $> 20$ times difference in $\dot M$ across the period gap.

The contrast in CN rate across the period gap does have some dependence on
$X_{^3\rm He}$.  Under gravitational braking only, as is the case below the
period gap, $\dot M$ is low enough that $M_{\rm ign}$ is insensitive to
$X_{^3\rm He}$.  However, the CN rate above the period gap, where magnetic
braking enhances $\dot M$, does depend on $X_{^3\rm He}$.  The dependence can
be estimated by using the ignition density, $\rho_{\rm ign}\propto
T^{-3.2}X_{^3\rm He}^{-1/2}$ from TB04, and combining with an ideal gas EOS,
$M_{\rm ign}\propto P_{\rm ign}\propto \rho_{\rm ign}T_{\rm ign}$, and the
power-law form of a free-free radiative envelope $T\propto P^{2/8.5}$, to
obtain $\nu_P\propto 1/M_{\rm ign}\propto X_{^3\rm He}^{0.33}$.  This scaling
is confirmed by numerical calculations.  Thus the expected maximum value of
$X_{^3\rm He}=0.003$ (see Section \ref{sec:CN}) leads to a 50\% enhancement
of the CN rate above the period gap over $X_{^3\rm He}=0.001$.  Matching this
new normal distribution to the observations at 3.5 hours, the rate below the
gap is 2/3 of that shown in figure \ref{fig:cumanddist}, still consistent
with the observations.  Thus the value we have adopted, $X_{^3\rm He}=0.001$
is expected from donor evolution and provides the best fit to the data when
birthrates above and below the period gap are assumed to be approximately
equal.

The observed orbital period distribution of CN is consistent with CVs
evolving across the period gap, and magnetic braking between 3 and 4
hours governed by a law like equation \ref{eq:mblaw}.  {\it As shown by the
thick (model) and thin (data) solid lines in figure \ref{fig:cumanddist}, the
contrast between the 1-2 and the 3-4 hour ranges is of the appropriate size
for a population in which $\dot J$ above 3 hours is enhanced by magnetic
braking.  This contrast in CN rate is entirely due to the dependence of
$M_{\rm ign}$ on $\dot M$.}

\subsection{Magnetic Accretors with No Magnetic Braking} 

The two components included in the model population, IMB and NMB, are also
shown separately as the dot-dashed and dotted lines.  These demonstrate the
significant enhancement that magnetic braking implies in the CN rate over
gravitational radiation alone.  The prediction of a NMB model with $b_P=0$
(all systems being born at long periods) falls weakly to shorter orbital
period in the normal distribution, as for the IMB curve below 2 hours.  With
a constant $b_P$, a NMB model monotonically increases to shorter orbital
periods as seen for the dotted lines.  Thus a NMB model alone is inconsistent
with the observed orbital period distribution.  The period distribution is
also satisfactorily reproduced without the addition of the NMB population,
making our confirmation of the IMB prescriptions and evolution across the gap
robust with some evidence for additional systems in and below the gap.
Measurement of this additional contribution is limited by the current quality
of data.

There are two reasons a CV might not undergo the MB phase.  For example, if
MB cannot work when the companion is fully convective, then there is a
minimum companion mass ($0.25M_\odot$) for having an MB phase.  We have
implemented this by placing all systems born with $P_{\rm orb}<2.5$ hours in
a NMB population.  This can also be seen in the the small number of systems
between 2 and 2.5 hours in the HNR population.  Additionally, motivated by
the lack of a clear period gap for magnetic CVs \citep{Verb97},
\citet{Lietal94} have conjectured that in Polars the WD magnetic field
prevents MB.  Recent $T_{\rm eff}$ measurements of magnetic WDs
\citep{Arauetal04,TownGans05} show that above the period gap, magnetic CVs
have a significantly lower $\dot M$ than Dwarf Novae at the same orbital
period, providing direct evidence that magnetic CV evolution is quite
different from that of non-magnetic systems.

The absolute relative populations of magnetic and nonmagnetic CVs is not well
known because they are dominated by different and difficult to quantify
selection effects.  Approximately 22\% of known CVs are magnetic
\citep{Arauetal04}.  Using this steady state two component population model
we find that, due to the slower evolution in the absence of MB, this requires
that 8\% of CVs born have strongly magnetic WDs, comparable to the fraction
among field WDs \citep{Jord97, WickFerr00}.  While giving tantalizing hints,
the small number of observed systems makes it impossible to separate the
prospective magnetic and non-magnetic contributions to the NMB population.

%%%%%%%%%%%%%%%%%%%%%%%%%%%%%%%%%%%%%%%%%%%%%%%%%%%%%%%%%%%%%%%%%%
\section{Discussion and Conclusion}
\label{sec:discussion}

We have shown that a very simple CV population model with the assumptions of
IMB and the $M_{\rm ign}$ calculated by TB04 reproduces the orbital period
distribution of CN,  supporting the idea that CVs evolve across the period
gap.  This agreement is independent of the outcome of the common envelope
phase or properties of the primordial binary population.
CN are
one CV phenomena that is sensitive to $\dot M$ averaged over the
longest timescales, the CN recurrence time, $10^5-10^7$ years.  The agreement
with IMB provides additional evidence that the IMB prescriptions do indeed
provide accurate long term average $\dot M$'s for CVs.  \citet{TownBild03}
similarly found that the WD $T_{\rm eff}$ in dwarf nova systems, which
reflects $\dot M$ due to compression of the envelope, matches the predictions
of IMB.  The existence of CN with $P_{\rm orb}$ in the period gap implies
that not all systems undergo IMB evolution, possibly due to inhibition of the
secondary's wind by a strong WD magnetic field \citep{Lietal94}.
This inhibition will lead to slower evolution of magnetic CVs, and thus
increase their relative number in the CV population.  We find that in order
to obtain the 22\% fraction of magnetic CVs observed \citep{Arauetal04}, a
relative birth fraction of magnetic CVs of 8\% is required, comparable to the
fraction among field WDs \citep{Jord97,WickFerr00}.

With a satisfactory $\dot M$ distribution in hand, the total number of
classical novae observed provides a measurement of the CV
population.  This conversion does not depend directly on CE
parameters, because a relatively small number of both observed novae
and CVs in general have $P_{\rm orb}> 4$ hours, where the birth
distribution is important.  The conversion does, however, depend on
the typical WD mass, due to its importance in setting $M_{\rm ign}$.
Note also that while most CV WDs have $M_{\rm WD}\lesssim0.7M_\odot$,
including many helium WDs (HNR), the average CN mass is sensitive to the
upper end of the mass distribution \citep{Nelsetal04}.
Knowledge of the $\dot M$ distribution allows proper adjustment for
the fact that while most CVs are below the period gap, most CN are in
CVs above the period gap.  This is a fairly short-lived period of the
CV's lifetime, lasting only $\sim 100$ Myr (HNR), which is preceded by
the post common envelope inspiral and followed by accretion below the
period gap during which the CN rate is much lower. {\it We find that
one CN per year implies $3\times 10^5$ ($9\times 10^5$) CVs above the
period minimum, $t\lesssim 3$ Gyr (4 Gyr) and a CV birthrate of $10^{-4}\
{\rm yr^{-1}}$ ($2\times 10^{-4}\ {\rm yr^{-1}}$) if the average CN
mass is $1.0 M_\odot$ ($0.6M_\odot$).}

Since pre-CVs finish the common envelope phase with a wide range of
orbital periods, the total CV and pre-CV population, and thus the CN
rate, should relate to the total amount of low-mass stars, which is
well represented by the K-band light. The apparent dependence of nova
rate on stellar population is still uncertain, but points to these
older populations.  Surveys of nova rates in external galaxies
\citep{WillShaf04} continue to show that the K-band specific nova rate
is about $2\pm 1\times 10^{-10}\ L_{\odot,K}^{-1}\ {\rm yr^{-1}}$ and
does not depend strongly on galaxy type in nearby galaxies, with the
exception of the LMC and SMC.  M31 shows evidence that the classical
nova rates might be dominated by the bulge population
\citep{ShafIrby01}.  Our work allows us to convert the K-band specific
Nova rate into the CV population and birthrate, giving 60-180 CVs
above the period minimum for every $10^6L_{\odot,K}$ in an old stellar
population and $2(4)\times 10^{-4}$ CVs born per year per
$10^{10}L_{\odot,K}$.  The 24-133 CVs estimated from {\it Chandra}
observations of 47 Tuc \citep{Heinetal05} is consistent with the
60-180 predicted by our work for the cluster mass ($10^6M_\odot$
\citealt{PryoMeyl93}), and agrees with the CN rate implied by the one
or two CNe observed in the last $\sim 140$ yr in the Galactic GCs
\citep{Sharetal04}.  Thus the GC CV population appears comparable to
that in the field.  The Type Ia supernova rate in elliptical galaxies
is $3.5^{+1.3}_{-1.1}\times 10^{-4}$ per year per $10^{10}L_{\odot,K}$
\citep{Mannetal04}, remarkably close to the CV birthrate. However, all
indications are that the WD masses in CVs are not nearly large enough
(TB03) to be Type Ia progenitors \citep{Branetal95}.

 The local galactic K-band luminosity density is 0.1-0.2 $L_{\odot,K}\
{\rm pc^{-3}}$ \citep{OlliMerr01}, giving a local CV space density of
approximately $9(27)\times 10^{-6}\ {\rm pc^{-3}}$ again for an
average mass of 1.0 (0.6) $M_\odot$.
This agrees well with the $2\times 10^{-5}\ {\rm pc^{-3}}$ found by
\citet{Poli96} from a theoretical calculation of the CV birthrate.
This space density is only slightly higher
than the $6\times 10^{-6}\ {\rm pc^{-3}}$ inferred from the PG survey
\citep{Ring96}, but more than 10 times the $8\times 10^{-7}$ pc$^{-3}$
obtained by \citet{Down86}. Both these survey numbers assume a scale
height of CVs of 150 pc, whereas the above K-band luminosity density
assumes a scale height of low mass stars of 300 pc.  Although more modern
color selected surveys exist, the lack of a broadly applicable, accurate CV
distance measure hinders improved measurements of the space density.

In a similar manner, the total rate of matter injected into the ISM
can be evaluated in terms of the CN outburst rate.  Since we are
assuming that all the matter transferred onto the WD is ejected in
each CN outburst, this is just the average $\dot M$ for the population
times the total number of systems.  We find that $3\times
10^{-5}M_\odot$ yr$^{-1}$ ($9\times 10^{-5}M_\odot$ yr$^{-1}$) is
ejected into the ISM for each CN/yr if the average CN mass is 1.0
$M_\odot$ ($0.6M_\odot$).  The average initial mass of the donor
contributes additional uncertainty, but generally less than that due
to the unknown average WD mass.  Though somewhat lower than the value
of $2\times 10^{-4} M_\odot$ per CN outburst found by
\citet{TrurLivi86}, these values are similar to the $\sim
10^{-4}M_\odot$ per outburst used by \citet{Gehretal98} to estimate
the impact of nucleosynthesis in CN outbursts on both the ISM and
primordial solar system material.  An interesting effect of the contrast
in $\dot M$ across the gap is that while approximately 85\% of CN
outbursts occur above the period gap and only 2\% of pre-period
minimum systems are above the gap, approximately equal processed
matter is contributed by objects above and below the period gap before
the period minimum.  This means that the type of CN which are observed
most often are only half the story for ISM injection. 

With proper calibration, classical novae provide a useful standard
candle for extragalactic distance measurements \citep{dellLivi95}.
The principle impediment to this usage is that the single measured
parameter available for calibration, the rate of decline, depends on
$M_{\rm WD}$, $\dot M$ and $T_c$.  TB04 calculated $T_c$ consistently
from $\dot M$, eliminating it as a free parameter.  Here we have
demonstrated that IMB provides a satisfactory prediction of the $\dot
M$ distribution in the Galactic CV population, finding that CN
outbursts occur predominantly when $\dot M\sim 10^{-9}M_\odot~{\rm
yr^{-1}}$, with a significant tail at lower $\dot M$.  This leaves
only $M_{\rm WD}$ to be calibrated from the rate of decline.  Thus,
with the support of theory, a statistically complete measurement of
nova magnitudes and decline rates may be sufficient to measure the
distance to an external galaxy.

We thank Boris G\"ansicke for discussion and early access to observational
data.  This work was supported by the National Science Foundation under
grants PHY99-07949, and AST02-05956, and by NASA through grant AR-09517.01-A
from STScI, which is operated by AURA, Inc., under NASA contract NAS5-26555.
DMT is supported by the NSF Physics Frontier Centers' Joint Institute for
Nuclear Astrophysics under grant PHY 02-16783 and DOE under grant DE-FG
02-91ER 40606.

%BIB

\bibliography{CN}

\begin{thebibliography}{52}
\expandafter\ifx\csname natexlab\endcsname\relax\def\natexlab#1{#1}\fi

\bibitem[{{Andronov} {et~al.}(2003){Andronov}, {Pinsonneault}, \&
  {Sills}}]{Andretal03}
{Andronov}, N., {Pinsonneault}, M., \& {Sills}, A. 2003, \apj, 582, 358

\bibitem[{{Araujo-Betancor} {et~al.}(2005){Araujo-Betancor}, {G\"ansicke},
  {Long}, Beuermann, Sion, {de Martino}, \& Szkody}]{Arauetal04}
{Araujo-Betancor}, S., {G\"ansicke}, B.~T., {Long}, S.~K., Beuermann, K., Sion,
  E.~M., {de Martino}, D., \& Szkody, P. 2005, \apj, in press
  (astro-ph/0412180)

\bibitem[{{Branch} {et~al.}(1995){Branch}, {Livio}, {Yungelson}, {Boffi}, \&
  {Baron}}]{Branetal95}
{Branch}, D., {Livio}, M., {Yungelson}, L.~R., {Boffi}, F.~R., \& {Baron}, E.
  1995, \pasp, 107, 1019

\bibitem[{{D'Antona} \& {Mazzitelli}(1982)}]{DAntMazz82}
{D'Antona}, F. \& {Mazzitelli}, I. 1982, \apj, 260, 722

\bibitem[{{della Valle} \& {Livio}(1995)}]{dellLivi95}
{della Valle}, M. \& {Livio}, M. 1995, \apj, 452, 704

\bibitem[{{Diaz} \& {Bruch}(1997)}]{DiazBruc97}
{Diaz}, M.~P. \& {Bruch}, A. 1997, \aap, 322, 807

\bibitem[{{Downes}(1986)}]{Down86}
{Downes}, R.~A. 1986, \apj, 307, 170

\bibitem[{{Faulkner}(1971)}]{Faul71}
{Faulkner}, J. 1971, \apjl, 170, L99

\bibitem[{{Fujimoto}(1982)}]{Fuji82}
{Fujimoto}, M.~Y. 1982, \apj, 257, 767

\bibitem[{G\"ansicke \& Townsley(2005)}]{TownGans05}
G\"ansicke, B.~T. \& Townsley, D.~M. 2005, in preparation

\bibitem[{{Gehrz} {et~al.}(1998){Gehrz}, {Truran}, {Williams}, \&
  {Starrfield}}]{Gehretal98}
{Gehrz}, R.~D., {Truran}, J.~W., {Williams}, R.~E., \& {Starrfield}, S. 1998,
  \pasp, 110, 3

\bibitem[{{Hameury} {et~al.}(1988){Hameury}, {King}, {Lasota}, \&
  {Ritter}}]{Hameetal88}
{Hameury}, J.~M., {King}, A.~R., {Lasota}, J.~P., \& {Ritter}, H. 1988, \mnras,
  231, 535

\bibitem[{Heinke {et~al.}(2005)}]{Heinetal05}
Heinke, C.~O. {et~al.} 2005, \apj, in press (astro-ph/05030132)

\bibitem[{{Howell} {et~al.}(2001){Howell}, {Nelson}, \&
  {Rappaport}}]{Howeetal01}
{Howell}, S.~B., {Nelson}, L.~A., \& {Rappaport}, S. 2001, \apj, 550, 897

\bibitem[{{Iben} \& {Tutukov}(1984)}]{IbenTutu84}
{Iben}, I. \& {Tutukov}, A.~V. 1984, \apj, 284, 719

\bibitem[{{Jordan}(1997)}]{Jord97}
{Jordan}, S. 1997, in ASSL Vol. 214: White dwarfs, ed. J.~Isern, M.~Hernanz, \&
  E.~{Garc\'ia-Berro} (Dordrecht: Klewer), 397

\bibitem[{{Kolb}(1993)}]{Kolb93}
{Kolb}, U. 1993, \aap, 271, 149

\bibitem[{{Kolb} \& {Baraffe}(1999)}]{KolbBara99}
{Kolb}, U. \& {Baraffe}, I. 1999, \mnras, 309, 1034

\bibitem[{{Li} {et~al.}(1994){Li}, {Wu}, \& {Wickramasinghe}}]{Lietal94}
{Li}, J.~K., {Wu}, K.~W., \& {Wickramasinghe}, D.~T. 1994, \mnras, 268, 61

\bibitem[{{MacDonald}(1984)}]{MacD84}
{MacDonald}, J. 1984, \apj, 283, 241

\bibitem[{Mannucci {et~al.}(2004)}]{Mannetal04}
Mannucci, F. {et~al.} 2004, \aap, in press (astro-ph/0411450)

\bibitem[{{McDermott} \& {Taam}(1989)}]{McDeTaam89}
{McDermott}, P.~N. \& {Taam}, R.~E. 1989, \apj, 342, 1019

\bibitem[{{Nelson} {et~al.}(2004){Nelson}, {MacCannell}, \&
  {Dubeau}}]{Nelsetal04}
{Nelson}, L.~A., {MacCannell}, K.~A., \& {Dubeau}, E. 2004, \apj, 602, 938

\bibitem[{{Nomoto}(1982)}]{Nomo82}
{Nomoto}, K. 1982, \apj, 253, 798

\bibitem[{{Olling} \& {Merrifield}(2001)}]{OlliMerr01}
{Olling}, R.~P. \& {Merrifield}, M.~R. 2001, \mnras, 326, 164

\bibitem[{{Paczy{\' n}ski}(1970)}]{Pacz70}
{Paczy{\' n}ski}, B. 1970, Acta Astronomica, 20, 47

\bibitem[{{Paczynski} \& {Sienkiewicz}(1981)}]{PaczSien81}
{Paczynski}, B. \& {Sienkiewicz}, R. 1981, \apjl, 248, L27

\bibitem[{{Paczynski} \& {Sienkiewicz}(1983)}]{PaczSien83}
---. 1983, \apj, 268, 825

\bibitem[{{Piersanti} {et~al.}(2000){Piersanti}, {Cassisi}, {Iben}, \&
  {Tornamb{\' e}}}]{Pieretal00}
{Piersanti}, L., {Cassisi}, S., {Iben}, I.~J., \& {Tornamb{\' e}}, A. 2000,
  \apj, 535, 932

\bibitem[{{Politano}(1996)}]{Poli96}
{Politano}, M. 1996, \apj, 465, 338

\bibitem[{{Prialnik} \& {Kovetz}(1995)}]{PriaKove95}
{Prialnik}, D. \& {Kovetz}, A. 1995, \apj, 445, 789

\bibitem[{{Pryor} \& {Meylan}(1993)}]{PryoMeyl93}
{Pryor}, C. \& {Meylan}, G. 1993, in Astronomical Society of the Pacific
  Conference Series, 357

\bibitem[{{Rappaport} {et~al.}(1982){Rappaport}, {Joss}, \&
  {Webbink}}]{Rappetal82}
{Rappaport}, S., {Joss}, P.~C., \& {Webbink}, R.~F. 1982, \apj, 254, 616

\bibitem[{{Rappaport} {et~al.}(1983){Rappaport}, {Verbunt}, \&
  {Joss}}]{Rappetal83}
{Rappaport}, S., {Verbunt}, F., \& {Joss}, P.~C. 1983, \apj, 275, 713

\bibitem[{{Ringwald}(1996)}]{Ring96}
{Ringwald}, F.~A. 1996, in ASSL Vol. 208: IAU Colloq. 158: Cataclysmic
  Variables and Related Objects, ed. A.~Evans \& J.~H. Wood (Dordrecht:
  Kluwer), 89

\bibitem[{Ritter \& Kolb(2003)}]{RittKolb03}
Ritter, H. \& Kolb, U. 2003, \aap, 404, 301, (update RKcat7.4)

\bibitem[{{Ritter} {et~al.}(1991){Ritter}, {Politano}, {Livio}, \&
  {Webbink}}]{Rittetal91}
{Ritter}, H., {Politano}, M., {Livio}, M., \& {Webbink}, R.~F. 1991, \apj, 376,
  177

\bibitem[{{Shafter}(1992)}]{Shaf92}
{Shafter}, A.~W. 1992, \apj, 394, 268

\bibitem[{{Shafter} \& {Irby}(2001)}]{ShafIrby01}
{Shafter}, A.~W. \& {Irby}, B.~K. 2001, \apj, 563, 749

\bibitem[{{Shara}(1980)}]{Shar80}
{Shara}, M.~M. 1980, \apj, 239, 581

\bibitem[{{Shara}(1989)}]{Shar89}
---. 1989, \pasp, 101, 5

\bibitem[{{Shara} {et~al.}(2004){Shara}, {Zurek}, {Baltz}, {Lauer}, \&
  {Silk}}]{Sharetal04}
{Shara}, M.~M., {Zurek}, D.~R., {Baltz}, E.~A., {Lauer}, T.~R., \& {Silk}, J.
  2004, \apjl, 605, L117

\bibitem[{{Spruit} \& {Ritter}(1983)}]{SpruRitt83}
{Spruit}, H.~C. \& {Ritter}, H. 1983, \aap, 124, 267

\bibitem[{{Townsley} \& {Bildsten}(2003)}]{TownBild03}
{Townsley}, D.~M. \& {Bildsten}, L. 2003, \apjl, 596, L227

\bibitem[{{Townsley} \& {Bildsten}(2004)}]{TownBild04}
---. 2004, \apj, 600, 390, {TB}

\bibitem[{{Truran} \& {Livio}(1986)}]{TrurLivi86}
{Truran}, J.~W. \& {Livio}, M. 1986, \apj, 308, 721

\bibitem[{{Verbunt}(1997)}]{Verb97}
{Verbunt}, F. 1997, \mnras, 290, L55

\bibitem[{{Verbunt} \& {Zwaan}(1981)}]{VerbZwaa81}
{Verbunt}, F. \& {Zwaan}, C. 1981, \aap, 100, L7

\bibitem[{{Warner}(1995)}]{Warn95}
{Warner}, B. 1995, {Cataclysmic Variable Stars} (Cambridge: Cambridge Univ.\
  Press)

\bibitem[{{Warner}(2002)}]{Warn02}
{Warner}, B. 2002, in AIP Conf. Proc. 637: Classical Nova Explosions, 3--15

\bibitem[{{Wickramasinghe} \& {Ferrario}(2000)}]{WickFerr00}
{Wickramasinghe}, D.~T. \& {Ferrario}, L. 2000, \pasp, 112, 873

\bibitem[{{Williams} \& {Shafter}(2004)}]{WillShaf04}
{Williams}, S.~J. \& {Shafter}, A.~W. 2004, \apj, 612, 867

\end{thebibliography}

\end{document}